\documentclass[letterpaper, 10 pt, conference]{ieeeconf}  

\IEEEoverridecommandlockouts                              

\usepackage{xcolor}
\usepackage{graphics} 
\usepackage{epsfig} 
\usepackage{mathptmx} 
\usepackage{times} 
\usepackage{amsmath} 
\usepackage{amssymb}  
\usepackage{subcaption}
\usepackage[switch]{lineno}
\usepackage{multirow}
\usepackage{makecell}
\usepackage{booktabs}
\usepackage{acro}
\usepackage{url}

\DeclareAcronym{DoF}{
short=DoF,
long=degree of freedom
}

\DeclareAcronym{NIR}{
short=NIR,
long=near-infrared
}

\DeclareAcronym{CAD}{
short=CAD,
long=solidworks
}

\DeclareAcronym{BCE}{
short=BCE,
long=binary cross entropy
}

\DeclareAcronym{MSE}{
short=MSE,
long=mean square error
}

\DeclareAcronym{DSC}{
short=DSC,
long=dice coefficient
}

\DeclareAcronym{CVBS}{
short=CVBS,
long=composite video broadcast signal
}

\DeclareAcronym{IVBag}{
short=IV bag,
long=intravenous bag
}

\DeclareAcronym{HDMI}{
short=HDMI,
long=high definition multimedia interface
}

\DeclareAcronym{GT}{
short=GT,
long=ground truth
}

\DeclareAcronym{DCNN}{
short=DCNN,
long=deep convolutional neural network
}

\DeclareAcronym{FCN}{
short=FCN,
long=fully convolutional network
}

\DeclareAcronym{HIV}{
short=HIV,
long=human immunodeficent virus
}

\title{\LARGE \bf VeniBot: Towards Autonomous Venipuncture with Automatic Puncture Area and Angle Regression from NIR Images}

\author{Xu Cao$^{1}$, Zijie Chen$^{1}$, Bolin Lai$^{2}$, Yuxuan Wang$^{1}$, Yu Chen$^{1}$,  Zhengqing Cao$^{1}$, Zhilin Yang$^{1}$, Nanyang Ye$^{3}$, \\ Junbo Zhao$^{4}$,  Xiao-Yun Zhou$^{5}$, Peng Qi$^{1*}$
\thanks{This work is supported by the National Natural Science Foundation of China (Number 51905379) and Shanghai Science and Technology Development Funds (Number 20QC1400900)}%
\thanks{$^{1}$ Xu Cao, Zijie Chen, Zhengqing Cao, Zhilin Yang, Yu Chen, Yuxuan Wang and Peng Qi are with Tongji University, Shanghai, China. {\tt\small pqi@tongji.edu.cn}}%
\thanks{$^{2}$ Bolin Lai is with PingAn Technology Co. Ltd., Shanghai, China.}%
\thanks{$^{3}$ Nanyang Ye is with Shanghai Jiao Tong University, Shanghai, China.}%
\thanks{$^{4}$ Junbo Zhao is with Zhejiang University, Hangzhou, China.}%
\thanks{$^{5}$ Xiao-Yun Zhou is with PAII Inc., MD, USA.}%
\thanks{$^{*}$ corresponding author.}%
}

\begin{document}
\maketitle
\thispagestyle{empty}
\pagestyle{empty}

\begin{abstract}
Venipucture is a common step in clinical scenarios, and is with highly practical value to be automated with robotics. Nowadays, only a few on-shelf robotic systems are developed, however, they can not fulfill practical usage due to varied reasons. In this paper, we develop a compact venipucture robot --- VeniBot, with four parts, six motors and two imaging devices. For the automation, we focus on the positioning part and propose a Dual-In-Dual-Out network based on two-step learning and two-task learning, which can achieve fully automatic regression of the suitable puncture area and angle from \ac{NIR} images. The regressed suitable puncture area and angle can further navigate the positioning part of VeniBot, which is an important step towards a fully autonomous venipucture robot. Validation on 30 VeniBot-collected volunteers shows a high mean \ac{DSC} of 0.7634 and a low angle error of $15.58^\circ$ on suitable puncture area and angle regression respectively, indicating its potentially wide and practical application in the future.
\end{abstract}

\section{INTRODUCTION}

Nowadays, venipuncture is the first step of clinical practice and has been exclusively performed in both routine examination and emergency. Statistics show that 1.4 billion times per year of venipuncture has been performed in the United States \cite{mccann2008continuing}. In the standard settings, after visually locating a proper superficial vein in the antecubital area, a needle is used to pierce the skin and introduced into the lumen parallelly to the longitudinal axis of vein. Once venous blood is extracted, the establishment of venous access is accomplished and further procedures become practicable \cite{campbell1999practical}. It is not a very complex procedure, however, reports have shown that the success rate of venipuncture could be $<50\%$ \cite{mbamalu1999methods}. Venipuncture is difficult even for professional clinicians, for elder people, infants, or people with shock state. Complication and infection also come along with venipuncture, for example, hematoma, phlebitis, artery or nerve damage \cite{stitik2001phlebotomy,simin2019incidence, lv2020incidence}, irborne diseases like corona virus \cite{scoppettuolo2020vascular} and hepatitis and human immunodeficent virus (HIV) \cite{mengistu2021worldwide}. Hence, venipucture is a procedure that takes time, manpower and experience, which is in an essential need to be automated.

A few robotic platforms have been developed, \textit{i.e.}, (1) Veebot, which includes a puncturing part and a robotic arm, is navigated by a infrared light and ultrasound, needs a medical staff to assist in attaching the appropriate test tube or \ac{IVBag} \cite{perry2013profile}; (2) Venouspro from Vasculogic, which is smaller and more portable than Veebot, includes a six \ac{DoF} positioning part and a three \ac{DoF} distal manipulator, is navigated by 3D \ac{NIR} and ultrasound \cite{chen2020deep}; (3) Anatomical structure tracking system, it controls a wireless ultrasound scanner to scan the vein with a robotic arm (KUKA LBR Med, KUKA AG, Germany) \cite{unger2020robot}.

In this paper, we develop a compact venipucture robot --- VeniBot. It includes a imaging, puncturing, positioning and supporting part. We also develop an automatic positioning algorithm to automatically find the suitable puncture areas and their longitudinal angle on the \ac{NIR} image.

\begin{figure}[ht]
    \centering
    \includegraphics[width=0.3\textwidth]{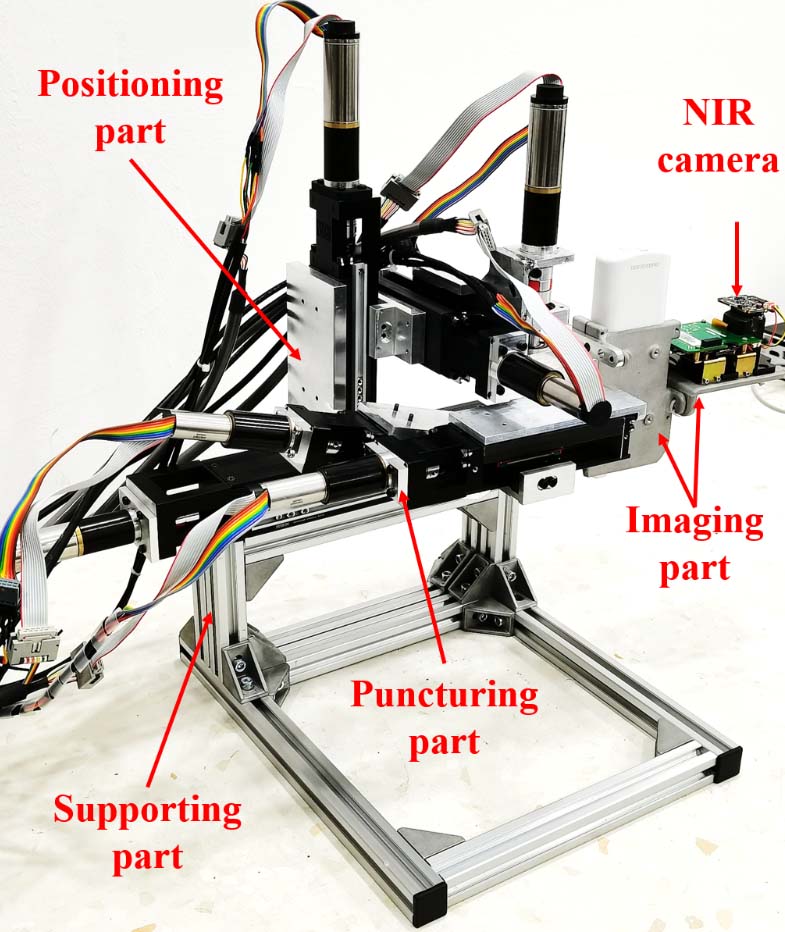}
    \caption{An illustration of the developed Venibot, with the imaging part, puncturing part, positioning part, supporting part and the \ac{NIR} camera.}
    \label{fig:real_venibot}
\end{figure}

For the problem of determining the suitable puncture area and angle from the \ac{NIR} image, traditional methods usually utilize features, \textit{i.e.}, the location, size and curvature of vein sections. However, it is inevitable for experts or experienced clinicians to build a specialized feature extractor and classifier according to clinical experience. Recently, the development of \ac{DCNN} can extract and classify the features automatically with multiple non-linear modules \cite{RN1}. Among \ac{DCNN} tasks, including image classification \cite{krizhevsky2012imagenet}, object detection \cite{girshick2015fast}, etc., segmentation approaches our aim most. Segmentation related \ac{DCNN} works at early stage usually take advantage of the sliding window-based method, \textit{i.e.}, Deepmedic \cite{2017Efficient}. However, the sliding window-based method wastes a lot of computational resources in repeatedly computing the network activation of overlapping regions between image parts. Hence, they were replaced later by \ac{FCN} \cite{2015Fully}. On medical image or volume, UNet \cite{ronneberger2015u} \cite{20163D}, which is a typical FCN, has shown outstanding performances and is used in this paper as the basic network structure.

Due to the fact that, suitable puncture area and angle determination is a highly abstract task, pure and simple segmentation can not achieve reasonable performance, as shown in Sec. \ref{sec:result}. In this paper, we propose a new Dual-In-Dual-Out network based on ResNeXt50-UNet, which fully explores the advantage of two-step learning and two-task learning, to fulfill the automatic puncture area and angle regression, hence to achieve an automatic positioning part.

In the following sections, we introduce the hardware design of VeniBot in Sec. \ref{sec:hardware} and the automatic algorithm of determining puncture area and angle in Sec. \ref{sec:puncture}, illustrate the experimental setup in Sec. \ref{sec:experiment} and show the results in Sec. \ref{sec:result}.

\section{METHODOLOGY}

\subsection{Hardware Design of VeniBot}
\label{sec:hardware}

\begin{figure}[ht]
    \centering
    \includegraphics[width=0.3\textwidth]{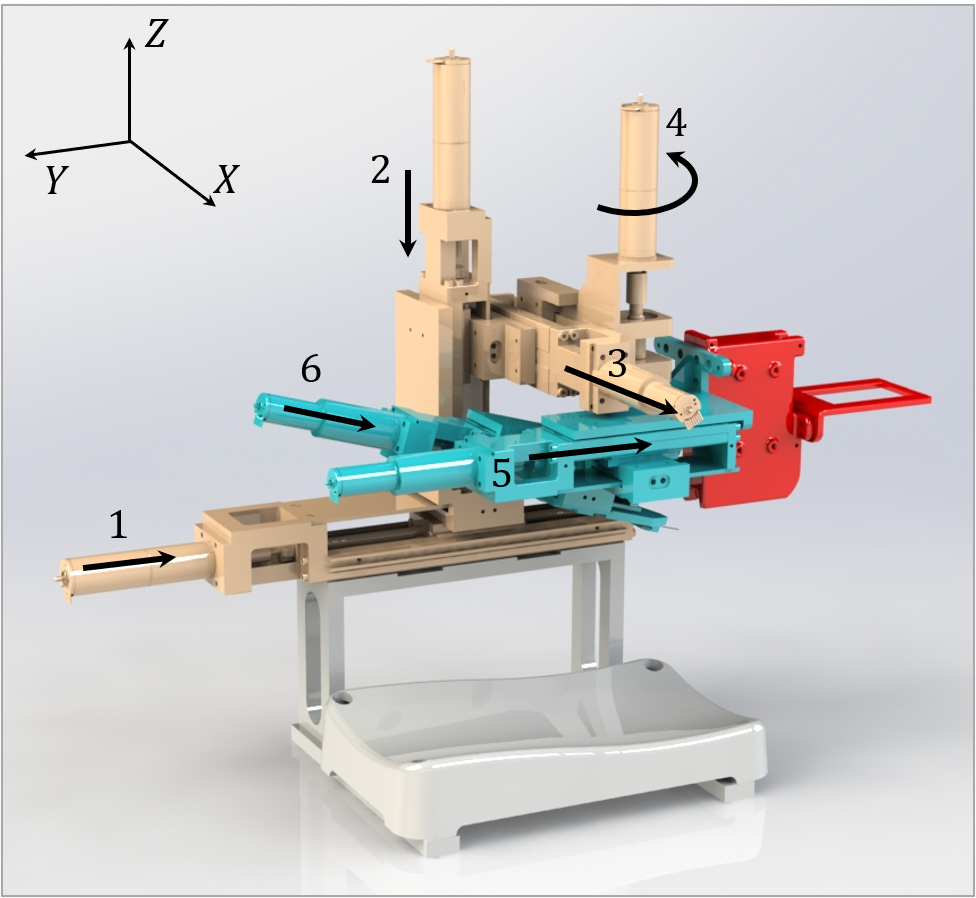}
    \caption{A visual illustration of the \ac{CAD} design of Venibot. The part in red/light blue/light brown/gray color is the imaging/puncturing/positioning/supporting part respectively.}
    \label{fig:cad_venibot}
\end{figure}

VeniBot includes four parts: (1) the imaging part which holds the \ac{NIR} and ultrasound device; (2) the puncturing part which accesses the vein; (3) the positioning part which delivers the puncturing part to suitable puncture areas; (4) the supporting part which holds the imaging, puncturing and positioning part. A visual illustration of the \ac{CAD} design of VeniBot is shown in Fig. \ref{fig:cad_venibot}. The workflow of VeniBot is that, first, the positioning part delivers the puncturing part to suitable puncture areas; second, the puncturing part accesses the vein. In this paper, we mainly focus on the automation of the positioning part while the automation of the puncturing part will be worked on in the other IROS submission.

\begin{figure}[ht]
    \centering
    \includegraphics[width=0.3\textwidth]{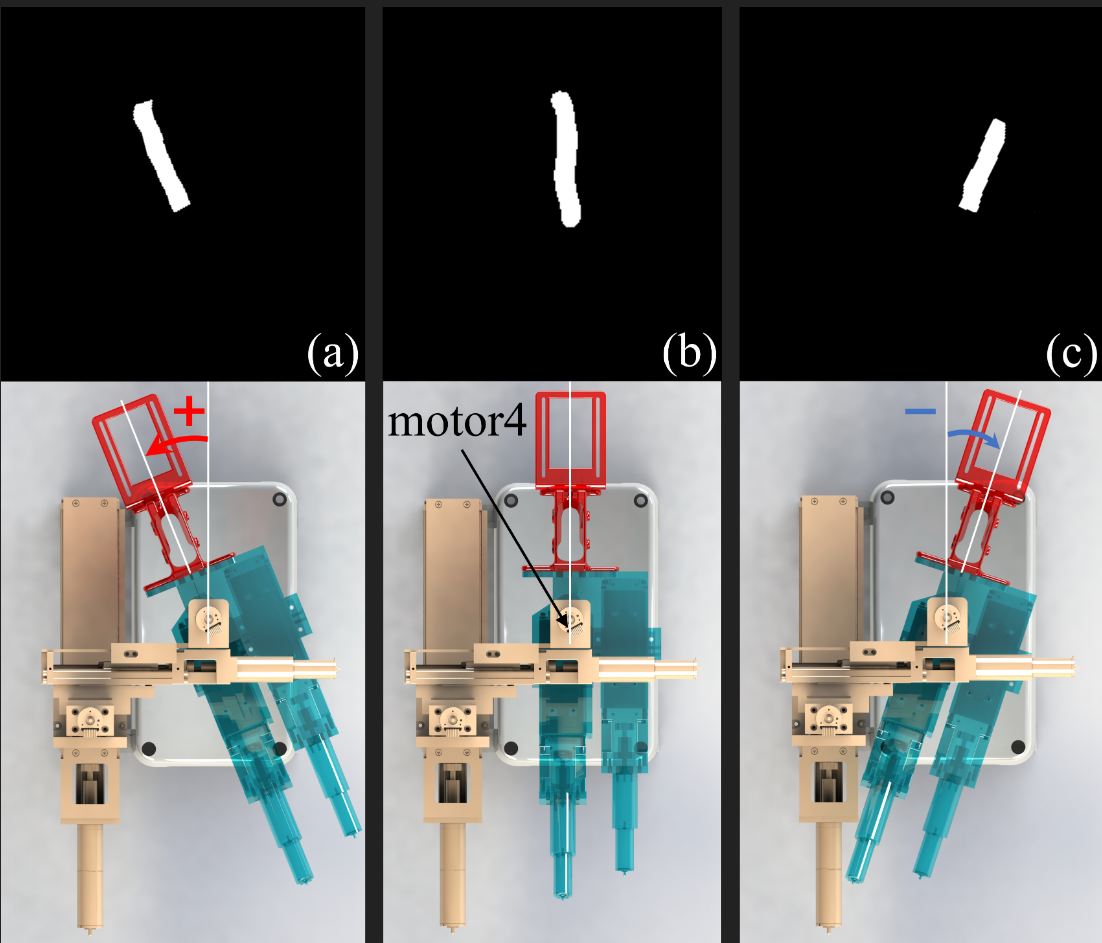}
    \caption{An illustration of the rotation of motor 4 along the z axis navigated by the regression of suitable puncture angle. (a/b) are with positive/negative angles.}
    \label{fig:motor4}
\end{figure}

On top of the supporting part, we assemble the positioning part which is with four \ac{DoF}s, as indicated in the light brown color in Fig. \ref{fig:cad_venibot}. The main function of positioning part is to deliver the puncturing part to suitable puncture areas, including the puncture position and angle in the xOy plane, under the navigation of \ac{NIR} images. Four motors consist the positioning part where three of them (motor 1, 2, and 3) are with one \ac{DoF} of translation along the y, z, and x axis respectively and one of them (motor 4) is with one \ac{DoF} of rotation along the z axis. A \ac{NIR} device (Projection Vein Finder VIVO500S from Shenzhen Vivolight Medical Device \& Technology Co., Ltd) is mounted at the base support on the imaging part. It scans the forearm of a volunteer, and from its \ac{NIR} image, the proposed Dual-In-Dual-Out network in Sec. \ref{sec:puncture} automatically regresses the suitable puncture area and angle in the xOy plane. Then motor 1 and 3 move along the y and x axis under the guidance of xOy coordinates and angle of the puncture area. As shown in Fig. \ref{fig:motor4}, Motor 4 rotates along the z axis under the guidance of puncture angle. After motor 1, 3, and 4 move to the target xOy position, motor 2 moves downward along the z axis, until the ultrasound probe mounted on the imaging part touches the volunteer's skin properly and hence a clear and normal ultrasound image can be collected for navigating the puncturing part further.

The bottleneck in the automation of positioning part is the determination of suitable puncture area and angle from the \ac{NIR} images, from which, VeniBot can deliver the puncturing part to the target site easily.

\begin{figure}
    \centering
    \includegraphics[width=0.5\textwidth]{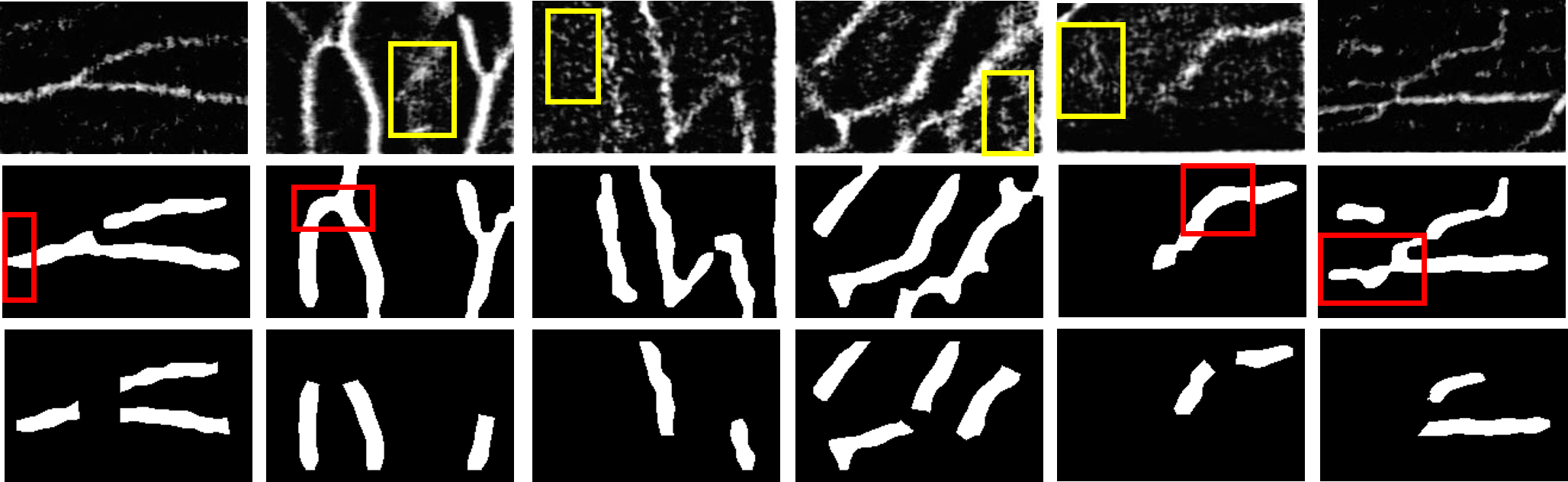}
    \caption{Six examples of vein segmentation and suitable puncture areas: (top) original image; (middle) vein segmentation \ac{GT}; (bottom) suitable puncture area \ac{GT}. Red boxes illustrate the unsuitable puncture areas while yellow boxes illustrate the noises.}
    \label{fig:puncture examples}
\end{figure}

\subsection{Suitable Puncture Area and Angle Regression}
\label{sec:puncture}

We randomly select six \ac{NIR} images as examples and show the vein segmentation and suitable puncture areas in Fig. \ref{fig:puncture examples}. First, different from the existing research on artery–vein segmentation in fundus images \cite{2019Artery}, the forearm vein segmentation faces noise caused by injury positions, dark skin blemishes, and hairs on the skin, as shown in the yellow boxes in Fig. \ref{fig:puncture examples}-top. Second, as shown in Fig. \ref{fig:puncture examples}-bottom, the suitable puncture area is a straight, large diameter and long venous area. While the red boxes in Fig. \ref{fig:puncture examples}-middle, vein bifurcations, large curvature vein and short vein are not suitable for puncture. Determining the suitable puncture area is very difficult, as it is too close to a segmentation task and distinguishing between segmenting all or part of the vein is difficult for a \ac{DCNN}. Moreover, to learn the abstract semantic information, such as the birfurcation and tortuosity of vein, is tough for a \ac{DCNN}.  This is illustrated in the results of traditional Single-In-Single-Out and Single-In-Dual-Out network in Sec. \ref{sec:result}, where both networks can not output suitable puncture areas properly.

Suitable puncture angle determination is even more difficult than the suitable puncture area determination, as it is more abstract and difficult for a \ac{DCNN} to learn. This is illustrated in Sec. \ref{sec:result} where a traditional Single-In-Single-Out network can not output any useful angle information.

\begin{figure}[ht]
    \centering
    \includegraphics[width=0.5\textwidth]{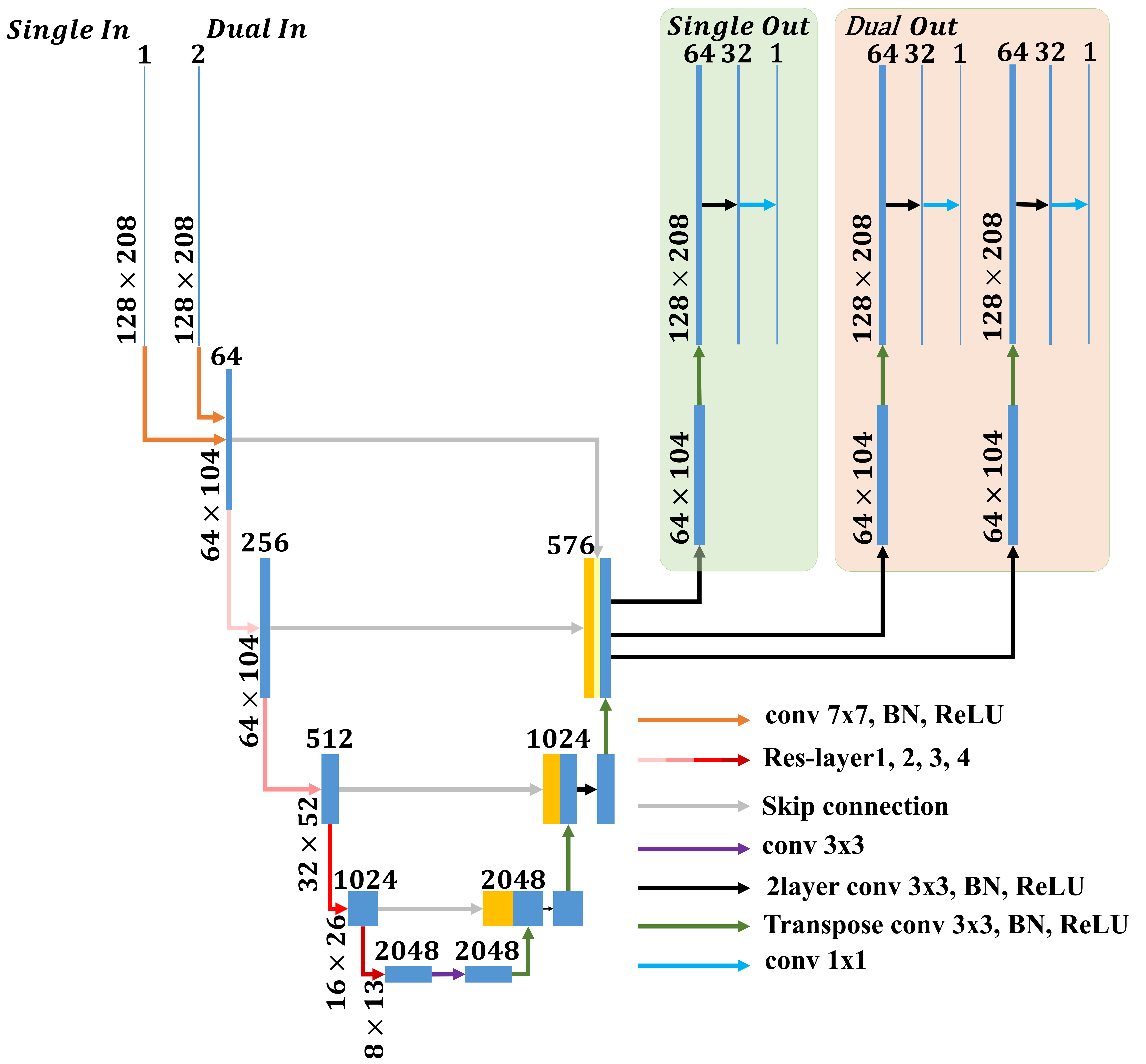}
    \caption{An illustration of the Dual-In-Dual-Out ResNeXt50-Unet network structure.}
    \label{fig:unet_resnext50}
\end{figure}

In this paper, we propose a Dual-In-Dual-Out network with two-step learning and two-task learning to determine the suitable puncture area and angle from the \ac{NIR} image inputs. A visual illustration of the proposed network is shown in Fig. \ref{fig:unet_resnext50}. It contains two steps of training: first, it trains a Single-In-Single-Out network to segment the vein from the \ac{NIR} image; second, it inputs both the \ac{NIR} image and the vein segmentation from the first step training into the Dual-In-Dual-Out network to regress the suitable puncture area and angle. Considering the population of UNet \cite{ronneberger2015u,zhou2019normalization} and ResNeXt\cite{2017Aggregated}, we choose the UNet model with the backbone of ResNeXt50 to fulfill the basic network architecture, details of each layer are shown in Tab. \ref{tab:network}. In the first step of training, \ac{BCE} loss is used as the loss function, while \ac{DSC} is used to evaluate the segmentation accuracy. Considering that our input image size is $128\times 208$ (208 is not divisible by 32), we omit the first Max-pooling layer in a standard ResNeXt50. We follow the idea of UNet - copy and crop both the first and second block to the corresponding up-sampling block.  In details, the encoder, which consists of multiple convolutional and down-sampling layers, can extract features at different scales from the input image. While the decoder, which consists of multiple  convolutional and up-sampling layers, can recover the size of feature map to the same size as the original input image.

\begin{table}[!htbp]
    \centering
    \caption{Details of each layer in the basic network structure - ResNeXt50-UNet.}
    \begin{tabular}{c|c|c}
    \hline
    stage&output&ResNeXt50-UNet\\ 
    \hline
    Conv0&$64\times104$&$7\times7$, 64, stride 2\\
    \hline
    Res-layer1&$64\times104$
    &$\left[\begin{array}{l}
         1\times1, 128  \\
         3\times3, 128, C=32\\
         1\times1, 256
    \end{array}\right]\times3$\\
    \hline
    Res-layer2&$32\times52$
    &$\left[\begin{array}{l}
         1\times1, 256  \\
         3\times3, 256, C=32\\
         1\times1, 512
    \end{array}\right]\times5$\\
    \hline
    Res-layer3&$16\times26$
    &$\left[\begin{array}{l}
         1\times1, 512  \\
         3\times3, 512, C=32\\
         1\times1, 1024
    \end{array}\right]\times6$\\
    \hline
    Res-layer4&$8\times13$
    &$\left[\begin{array}{l}
         1\times1, 1024  \\
         3\times3, 1024, C=32\\
         1\times1, 2048
    \end{array}\right]\times3$\\
    \hline
    Conv5&$8\times13$&$3\times3$, 2048, stride 1\\
    \hline
    TransConv6&$16\times26$&$3\times3$, 1024, stride 2\\
    \hline
    \multirow{3}*{ConvBlock7}&
    \multirow{3}*{$16\times26$}&concatenate\\\cline{3-2}
    &&$3\times3$, 1024, stride 1\\\cline{3-2}
    &&$3\times3$, 1024, stride 1\\
    \hline
    
    TransConv8&$32\times52$&$3\times3$, 512, stride 2\\
    \hline
    \multirow{3}*{ConvBlock9}&
    \multirow{3}*{$32\times52$}&concatenate\\\cline{3-2}
    &&$3\times3$, 512, stride 1\\\cline{3-2}
    &&$3\times3$, 512, stride 1\\
    \hline

    TransConv10&$64\times104$&$3\times3$, 256, stride 2\\
    \hline
    \multirow{3}*{ConvBlock11}&
    \multirow{3}*{$64\times104$}&concatenate\\\cline{3-2}
    &&$3\times3$, 256, stride 1\\\cline{3-2}
    &&$3\times3$, 256, stride 1\\
    \hline
    
    TransConv12&$128\times208$&$3\times3$, 128, stride 2\\
    \hline
    
    \multirow{2}*{ConvBlock13}&
    \multirow{2}*{$128\times208$}&$3\times3$, 64, stride 1\\\cline{3-2}
    &&$3\times3$, 64, stride 1\\
    \hline

    Conv14&$128\times208$&$1\times1$, 1, stride 1\\
    
    \hline
    \multicolumn{2}{c|}{\#params}& $123.2\times10^{6}$ \\
    \hline
    \multicolumn{2}{c|}{FLOPs}& $52.1\times10^{9}$\\
    \hline
    \end{tabular}
    \label{tab:network}
\end{table}

In the second step of training, we use regression instead of segmentation to deal with the suitable puncture area and angle determination, as the segmentation strategy always outputs the concrete vein segmentation rather than abstract puncture area and angle prediction. For the output feature map, pixels' intensities are normalized into the $[0, 1]$ interval. We binary the output of ResNeXt50-UNet network with choosing the threshold based on experience. L2 distance is used as the loss fuction. \ac{DSC} is also used as the evaluation for the suitable puncture area 
regression. Except the proposed Dual-In-Dual-Out network, we also tried a few traditional methods, we list the details of each as below:

\subsubsection{Single-In-Single-Out network}

To predict the suitable puncture area and angle, the most common idea is to train a ResNeXt50-UNet to output the suitable puncture area and angle directly. However, this Single-In-Single-Out network can not perform well on the suitable puncture area regression and fail totally on the sutiable puncture angle regression in our experiments, because the feature of 'proper for puncture' is too abstract for a Single-In-Single-Out network to learn. So we turn to learn abstract features though two-step learning and two-task learning.

\subsubsection{Single-In-Dual-Out network}
We found that directly train a Single-In-Single-Out network to output the suitable puncture angle may lead to the non-convergence of the network. Also, we found that the Single-In-Single-Out network performs well in segmenting the vessel from the \ac{NIR} image. Considering that there may be a strong correlation between vessel segmentation and suitable puncture area/angle regression, we try to let the network both output the segmentation result and the puncture area/angle regression. We hope that this two-task learning with shared variables for dual outputs can help the network convergence and ensure a promising result of puncture area/angle regression.

We try to modify the model on the Single-In-Single-Out network. At first, we only change the last layer from one output to two outputs. However, too many shared variables cause the two outputs to be indistinguishable. Hence, we modify the scheme so that the model is divided into two parts from the last third block. The result of Single-In-Dual-Out network is better than the Single-In-Single-Out network.

\subsubsection{Dual-In-Single-Out network}
The two-task learning achieves some improvements. However, the accuracy is not high enough for medical usage. We further include two-step learning from concrete to abstract, and design a Dual-In-Single-Out network. One input is the ResNeXt50-UNet output of the vein segmentation before the sigmoid layer, which provides the information of vessel. Another is the original input image, providing complete information which may be lost during the vein segmentation. This Dual-In-Single-Out network shows a competitive result.

\subsubsection{Dual-In-Dual-Out network}
Inspired by the success of two-step learning and two-task learning, we further propose a Dual-In-Dual-Out network, where both the original image and the vein segmentation are used as the input. To save the training times, we set the two-task as suitable puncture area and angle regression respectively. Hence, train one Dual-In-Dual-Out network regresses both the suitable puncture area and angle.
\section{Experiment}
\label{sec:experiment}

\subsection{Dataset}
\begin{figure}[ht]
    \centering
    \includegraphics[width=0.3\textwidth]{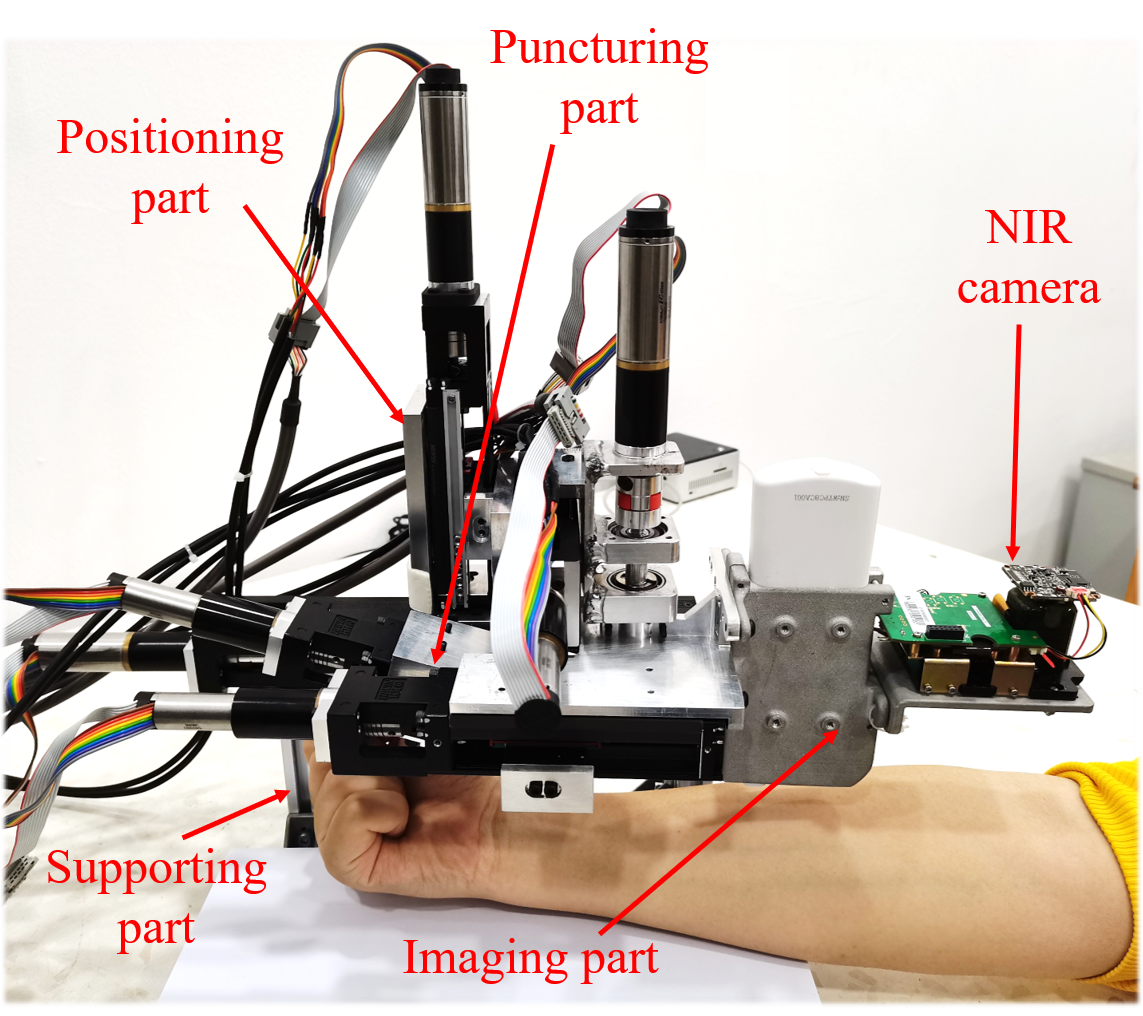}
    \caption{A demonstration of the data collection process with VeniBot and NIR camera.}
    \label{fig:dataset_collect}
\end{figure}

To verify the performance of the proposed VeniBot and Dual-In-Dual-Out network, we conduct experiments on 30 volunteers, including 14 females and 16 males, and with the age between 18-60 years old. All the images are collected by the proposed VeniBot and the \ac{NIR} device on it. The data collection process is demonstrated in Fig. \ref{fig:dataset_collect}. In total, we collected 900 \ac{NIR} images, with 30 images of the forearm vein per volunteer. Because the device is not equipped with a memory card, we added a \ac{CVBS} video transmission connector on the \ac{NIR} device and established a connection through \ac{CVBS} to \ac{HDMI} signal conversion \footnote{Due to the limitation of the design of \ac{NIR} device itself, \ac{CVBS} connector is the only choice.}, hence realizing the collection and storage of \ac{NIR} vein images.

The \ac{GT} of vein segmentation was labeled by self-defined labeling pipeline which mainly consists of operations, such as Gaussian filtering, eroding-dilating, brightness adjustment, histogram normalization, Hessian feature detection and binarization etc.. By manually adjusting the parameters image by image, the optimal segmentation \ac{GT} of each \ac{NIR} image is obtained. The suitable puncture area \ac{GT} was labeled manually by erasing areas that are not suitable for puncture, such as vein bifurcations, large curvature vein sections, veins sections close to the imaging edge of \ac{NIR} camera, and short vein sections.

\begin{figure}[ht]
    \centering
    \includegraphics[width=0.3\textwidth]{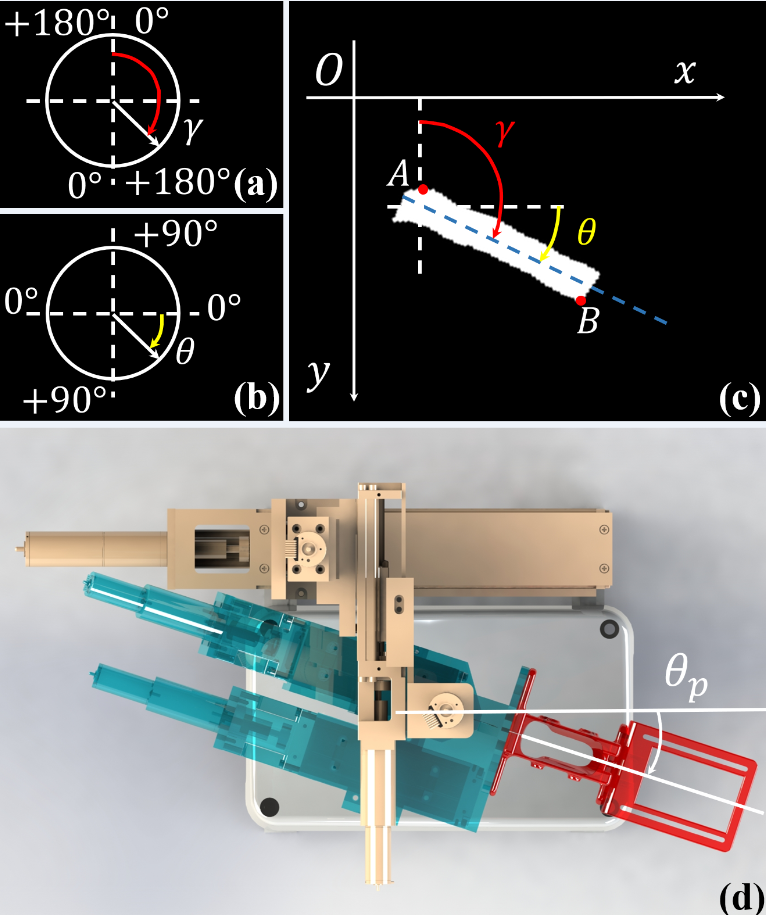}
    \caption{An illustration of the definition of \ac{GT} angle: (a) discontinuous angles of vein section; (b) continuous angles of vein section; (c) relationship between OpenCV angle and \ac{GT} angle; (d) the target angular position($\theta_p$) of puncturing part.}
    \label{fig:label_ang}
\end{figure}

Since the suitable puncture areas are straight and without obvious curvature or vein bifurcation, generally they are spindle-shaped. We performed ellipse fitting function of OpenCV on each vein section and got the corresponding angle.\footnote{Function $cv2.fitEllipse()$ returns the length of the major and minor axis of the ellipse and the angle of the longitudinal axis.} As shown in Fig. \ref{fig:label_ang}(b), the angle value is not continuous at the junction of $0^{\circ}$ and $180^{\circ}$, which may cause unstable training of neural networks. Hence, we calculate continuous angles by:
 
\begin{equation}
    \theta = |\gamma - 90^{\circ}|,
\end{equation}

where $\gamma$ is the longitudinal axis angle of the fitted ellipse, $\theta$ is the continuous angle and is shown in Fig. \ref{fig:label_ang}. In addition, $\theta$ can not distinguish between the clockwise and counterclockwise angle relative to the x-axis. Hence the difference between y-value of point A and B is used to determine the clockwise and counterclockwise of $\theta$.

In order to avoid over-fitting and to promote the generalization ability of the model, we propose to regularize the networks by adopting rich data augmentations, mainly including two kinds, that is, spatial data augmentation and intensity data augmentation. Spatial data augmentation includes random resize and crop, random horizontal and vertical flip, random rotation. Intensity data augmentation includes random brightness, contrast and saturation adjustments within the specified range. Details are listed in Tab. \ref{tab:augmentation}.

\begin{table}[!htbp]
\caption{The setup of spatial and intensity data augmentation used on vein segmentation model training and suitable puncture area/angle regression model training.}
\centering
\begin{tabular}{c|c|c|c}
\hline
\multicolumn{3}{c|}{data augmentation}&\makecell{variation range\\ or probability}\\

\hline
\multirow{5}{*}{spatial}
&\multicolumn{2}{c|}{random vertical flip}&50\% \\
\cline{2-4}
&\multicolumn{2}{c|}{random horizontal flip}&50\% \\
\cline{2-4}

&\multirow{3}{*}{\makecell{random crop \\and resize}}
&output size& $128\times208$\\
\cline{3-4}
& &crop scale& 0.2-1.0\\
\cline{3-4}
& &aspect ratio& 0.5-2\\
\cline{2-4}
&\multicolumn{2}{c|}{random rotation}&$-45^{\circ}\sim45^{\circ}$\\

\hline

\multirow{3}{*}{intensity }
&\multicolumn{2}{c|}{brightness}& 0.8$\sim$1.2\\
\cline{2-4}
&\multicolumn{2}{c|}{contrast}&$0.8\sim1.2$\\
\cline{2-4}
&\multicolumn{2}{c|}{saturation}&$0.8\sim1.2$\\
\hline
\end{tabular}
\label{tab:augmentation}
\end{table}

\subsection{Implementation details}

Including baselines, all experiments were carried out with a Win10(64-bit) computer, which is equipped with Intel i7-8750 CPU @2.2 GHz, 16 GB DDR4 memory and 4GB NVIDIA GTX 1050Ti GPU. 

All networks are built based on Pytorch. Adam \cite{2014Adam} with the weight decay as $10^{-5}$ was adopted to optimize the model. The batch size is 2. The epoch is 5, that is, 1575 iterations. The training process took around 1 hour on a NVIDIA GTX 1050Ti GPU. The learning rate was set as $10^{-3}$ at the beginning of training. Besides, during the experiments, we found that iterations required for model convergence was generally lower than 1000, so the validation interval in the training stage was selected as relative small, that is, 20 iterations in the second step training of Dual-In-Single-Out and Dual-In-Dual-Out network and 50 iterations in all other model training. In addition, ReduceLROnPlateau was adopted to schedule the learning rate with the decay factor as $0.5$ and patience as $5$. For the vein segmentation, the learning rate was scheduled according to the \ac{DSC} of each validation. While for suitable puncture area and angle regression, the learning rate was scheduled according to the \ac{MSE} of each validation.

\begin{table*}[h]
\centering
\caption{The mean and std \ac{DSC} (in percentage) of vein segmentation achieved by Single-In-Single-Out network, and that of suitable puncture area regression achieved by Single-In-Single-Out, Single-In-Dual-Out, Dual-In-Single-Out, and Dual-In-Dual-Out network.}
\begin{tabular}{c|c|cccccc}
\hline
Task & Method & Fold 0 & Fold 1 & Fold 2 & Fold 3 & Fold 4 & Average \\

\hline
Vein segmentation & Single-In-Single-Out & 77.34$\pm$10.56 & 76.93$\pm$12.47 & 73.19$\pm$10.17 & 79.26$\pm$6.99 & 76.76$\pm$8.36 & 76.70$\pm$10.09 \\

\hline
\multirow{4}{*}{\makecell[c]{Puncture regression}} & Single-In-Single-Out & 53.15$\pm$19.92 & 57.16$\pm$16.91 & 56.23$\pm$16.47 & 55.13$\pm$19.07 & 57.49$\pm$16.50 & 55.83$\pm$17.90  \\

& Single-In-Dual-Out & 54.77$\pm$19.73 & 54.85$\pm$18.11 & 56.67$\pm$16.95 & 56.66$\pm$18.93 & 57.39$\pm$17.78 & 56.80$\pm$18.36 \\

& \makecell{Dual-In-Single-Out} & 72.76$\pm$15.62
& \textbf{75.76$\pm$17.65}
&74.15$\pm$13.36 
&\textbf{77.02$\pm$18.90}
&\textbf{76.79$\pm$14.67}
&75.53$\pm$16.24 \\

& \makecell{Dual-In-Dual-Out} & \textbf{75.50$\pm$14.53} & 
75.30$\pm$17.87&
\textbf{78.00$\pm$13.85} &
74.15$\pm$13.36&
76.19$\pm$19.72
& \textbf{76.34$\pm$16.69} \\
\hline
\end{tabular}
\label{tab:result_dsc}
\end{table*}

\begin{figure}[!htbp]
    \centering
    \includegraphics[width=0.5\textwidth]{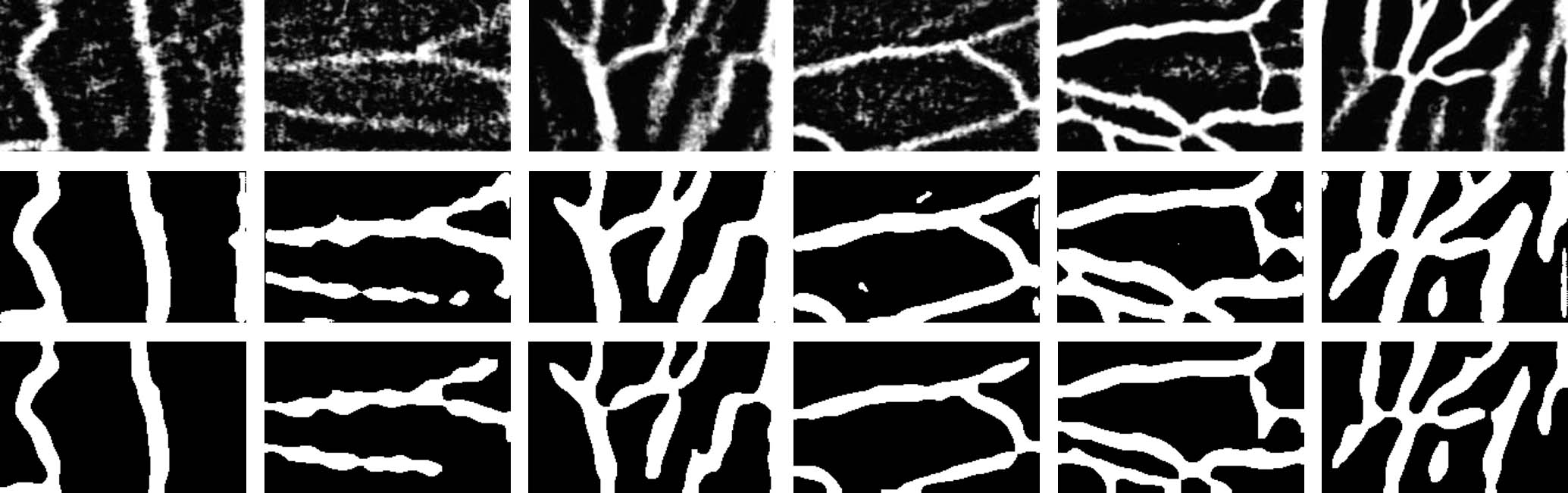}
    \caption{An illustration of six vein segmentation results: (top) original image; (middle) segmentation result of Single-In-Single-Out network; (bottom) \ac{GT}.}
    \label{fig:result_seg}
\end{figure}

\begin{figure}[htb]
    \centering
    \includegraphics[width=0.5\textwidth]{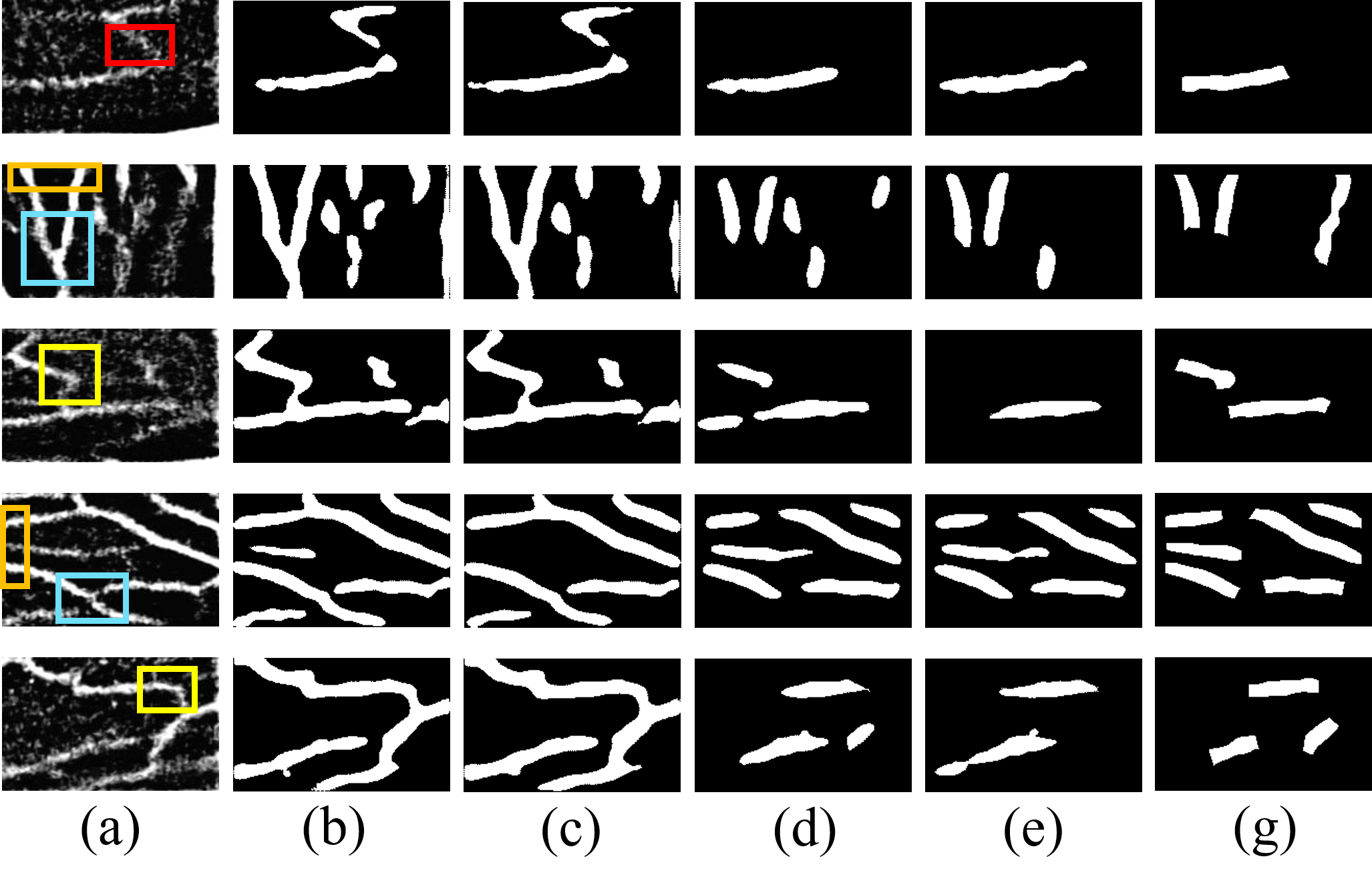}
    \caption{An illustration of five suitable puncture area regression results: (a) original image; (b/c/d/e) suitable puncture area regression results of Single-In-Single-Out/Single-In-Dual-Out/Dual-In-Single-Out/Dual-In-Dual-Out network; (g) suitable puncture area \ac{GT}. The red, orange, blue, and yellow box indicates non-suitable puncture areas, including the short vein, vein closed to the imaging edge of \ac{NIR} camera, vein bifurcation and largely curved vein respectively}
    \label{fig:result_pun}
\end{figure}

\begin{table*}[h]
\centering
\setlength{\tabcolsep}{2mm}
\caption{The mean and std angle error (in $^\circ$) of suitable puncture angle regression achieved by Single-In-Single-Out, Single-In-Dual-Out, Dual-In-Single-Out, and Dual-In-Dual-Out network.}
\begin{tabular}{c|c|cccccc}

\hline
Task & Method & Fold 0 & Fold 1 & Fold 2 & Fold 3 & Fold 4 & Average \\

\hline
\multirow{4}{*}{\makecell[c]{\makecell{Angle regression}}} & \makecell{Single-In-Single-Out} & - & - & - & - & - & - \\

& \makecell{Single-In-Dual-Out} & 20.07$\pm$16.91 & 25.32$\pm$20.13 & 25.25$\pm$21.61 & 23.08$\pm$20.02 & 23.89$\pm$18.86 & 23.57$\pm$19.68 \\

& Dual-In-Single-Out & 14.62$\pm$14.78 & 18.31$\pm$14.23 & \textbf{21.19$\pm$17.70} & 13.57$\pm$15.22 & 17.93$\pm$16.72 & 17.28$\pm$16.06  \\

& Dual-In-Dual-Out & \textbf{14.35$\pm$15.82} & \textbf{13.21$\pm$14.76} & 21.57$\pm$19.47 & \textbf{13.29$\pm$16.42} & \textbf{14.90$\pm$13.75} & \textbf{15.58$\pm$16.47}  \\
\hline

\end{tabular}
\label{tab:result_ang}
\end{table*}

\section{Result}
\label{sec:result}
Three stages of results are offered, including the vein segmentation, suitable puncture area regression and suitable puncture angle regression. Among these three stages, the suitable puncture area and angle regression is what VeniBot can directly been navigated by, while the vein segmentation is only used as a supplemental input to improve the performance of suitable puncture area and angle regression. Hence, Single-In-Single-Out, Single-In-Dual-Out, Dual-In-Single-Out, and Dual-In-Dual-Out are compared for the suitable puncture area and angle prediction, while only Single-In-Single-Out network is used on the vein segmentation.

\subsection{Vein Segmentation}
\label{sec:result_segmentation}
For the segmentation of vein from \ac{NIR} image, a Single-In-Single-Out network is used. The mean and std \ac{DSC} and several visual segmentation results are shown in Tab. \ref{tab:result_dsc} and Fig. \ref{fig:result_seg}, respectively. We can see that, even with only a simple Single-In-Single-Out network \footnote{The vein segmentation is a very basic and simple task, and is much easier than suitable puncture area and angle regression.}, a high mean \ac{DSC} of 0.767 and reasonable vein segmentation results are achieved. This vein segmentation result is used as an additional input to improve the following suitable puncture area and angle regression.

\subsection{Suitable Puncture Area Regression}
\label{sec:result_regression}
For the regression of suitable puncture areas, Single-In-Single-Out, Single-In-Dual-Out, Dual-In-Single-Out, and Dual-In-Dual-Out network are used. The mean and std \ac{DSC} are shown in Tab. \ref{tab:result_dsc}. For Single-In-Single-Out vs. Dual-In-Single-Out, and Single-In-Dual-Out vs. Dual-In-Dual-Out, we can see that, adding the vein segmentation as an additional input improves the performance of suitable puncture area regression significantly by almost 0.2 \ac{DSC}. While for Single-In-Single-Out vs. Single-In-Dual-Out, and Dual-In-Single-Out vs. Dual-In-Dual-Out, we can see that, adding the vein segmentation or suitable puncture angle regression as an additional output improves the performance of suitable puncture area regression slightly by almost 0.01 \ac{DSC}.

Five examples of the suitable puncture area regression by the four methods are shown in Fig. \ref{fig:result_pun}. We can visually see that both the Dual-In-Single-Out and Dual-In-Dual-Out network can distinguish between the suitable and non-suitable puncture area better than the Single-In-Single-Out and Single-In-Dual-Out network, indicating the importance and value of bringing the vein segmentation into the network's input.

\subsection{Suitable Puncture Angle Regression}
For the regression of suitable puncture angle, Single-In-Single-Out, Single-In-Dual-Out, Dual-In-Single-Out, and Dual-In-Dual-Out network are used. The mean and std angle error are shown in Tab. \ref{tab:result_ang}. Among the four methods, the Single-In-Single-Out network fails to predict any useful angle information. From the results of Single-In-Dual-Out, Dual-In-Single-Out, and Dual-In-Dual-Out network, we can see that, similar to the trend in suitable puncture area regression, both the Dual-Input and Dual-Output strategy improve the performance of suitable puncture angle regression notably.

Overall, these results indicate the great value and advantage of the two-step learning and two-task learning in the proposed Dual-In-Dual-Out network.
\section{DISCUSSION AND CONCLUSION}

In this paper, we build a compact venipuncture robot - VeniBot, and propose a novel Dual-In-Dual-Out network for suitable puncture area and angle regression. The Dual-In-Dual-Out network builds a pipeline that includes two models, that is, vein segmentation and puncture area/angle regression. It enables the end-to-end determination of vein, puncture area, and puncture angle simultaneously. We evaluate it on a newly VeniBot-collected \ac{NIR} dataset and it outperforms the other three baselines with remarkable margins. This paper focuses on the automation of positioning part, while the automation of puncturing part is introduced in another submission. In the future, we will work on integrating these two parts, formulating a fully-automatic VeniBot system.

\clearpage

\bibliographystyle{IEEEtran}
\bibliography{IROS_cx.bib}
\end{document}